# Impacts of Jet Stream Structure on Cyclone Merging and Persistent Anticyclones: Insights from Dry Idealized Simulations

Mingfei Ren[1], Gan Zhang[1], Kai-Yuan Cheng[2], Lucas Harris[3], Talia Tamarin-Brodsky[4], and Joseph Mouallem[2]

[1]Department of Climate, Meteorology & Atmospheric Sciences, University of Illinois Urbana-Champaign, Urbana, IL, USA.

[2]Atmospheric & Oceanic Sciences, Princeton University, Princeton, NJ, USA.

[3]GFDL/NOAA, Princeton, NJ, USA.

[4]Department of Earth, Atmospheric, and Planetary Sciences, Massachusetts Institute of Technology, Cambridge, MA, USA.

Corresponding author: Gan Zhang (gzhang13@illinois.edu)

**Key Points:**

- Idealized atmospheric simulations are conducted to explore how the jet stream structure influences baroclinic wave development.

- Broad and deep jet streams shifting poleward increases the merging of surface cyclones and contribute to wind extremes.

- Broad and deep jet stream also favors the development of surface high-pressure systems that stall for days.

**Abstract**

Midlatitude jet streams exhibit substantial variability in latitude, width, and vertical depth on synoptic to multi-decadal timescales. While the upper-level dynamics of baroclinic waves have been extensively studied, the sensitivity of the extreme-generating, low-level phenomena to these variations remains underexplored. Here, we systematically investigate this sensitivity using dry, adiabatic idealized experiments with the GFDL FV3 dry dynamical core initialized with analytically specified jets. We identify jet variations that control synoptic-scale features of interest. Results indicate that poleward-shifted jets accelerate initial cyclone intensification and favor anticyclonic Rossby Wave Breaking (RWB). These wave-breaking tendencies are consistent with established baroclinic paradigms, validating the newly configured idealized simulations. Additionally, jet width regulates the likelihood of surface cyclone merging. Poleward-shifted, broader, and higher jets produce more frequent cyclone merging, generating intense wind extremes. Finally, we show that poleward-shifted, broad, deep jets dynamically precondition the flow for persistent stationary anticyclones in the absence of diabatic contributions. Together, these findings illustrate how changes in jet stream structure may modulate midlatitude weather extremes.

**Plain Language Summary**

High-altitude bands of fast-moving air, known as jet streams, steer daily weather systems around the globe. These jets vary in shape across seasons and years, shifting poleward or equatorward and expanding in width or vertical depth. Although such natural fluctuations are well documented, it remains unclear how specific changes in jet-stream structure influence extreme surface weather. In this study, we use a simplified atmospheric model to isolate how different jet structures affect weather events. We find that when the jet shifts poleward, broadens, and deepens vertically, surface weather behavior changes markedly. Under these conditions, cyclones are more likely to collide and merge, producing brief bursts of extreme winds. The same broad, deep jet configurations also promote the stalling of high-pressure systems, which can persist for days and are often associated with prolonged heatwaves or cold spells. These results suggest that accurately capturing extreme weather risks requires a comprehensive understanding of how the jet stream's structural evolution—across seasonal and interannual scales—governs the dynamic coupling between high-altitude atmospheric states and surface weather extremes.

**1 Introduction**

The midlatitude jet stream acts as the primary waveguide for synoptic-scale weather systems, organizing the storm tracks that transport heat, momentum, and moisture across the hemispheres. Consequently, the variability of the jet stream is a central subject of dynamic meteorology, particularly regarding its response to internal atmospheric variability (Visbeck et al., 2001; Hurrell et al., 2003) and external forcings (Yin, 2005; Barnes & Polvani, 2013; Blackport & Fyfe, 2022; Perez et al. 2024). While the midlatitude jet shows robust changes (e.g., a poleward shift and a greater vertical depth) under greenhouse gas forcing in climate model simulations (Tan et al., 2019; Perez et al. 2024; Breul et al. 2025), the jet exhibits profound natural variability in its geometric structure on synoptic to interannual timescales. Observational analyses reveal that the jet undergoes continuous variations between poleward and equatorward displacements and between sharpening and widening phases (Woollings et al., 2018). Disentangling the independent impacts of these structural changes—specifically jet width and vertical depth—from latitudinal shifts is critical, as these geometric variations may fundamentally precondition the atmosphere for compound extreme weather events.

Classical theories of baroclinic instability provide a foundational understanding of how the jet stream initiates weather systems. The linear relationships between jet speed, vertical shear, and the growth rate of baroclinic waves are well-established (Charney, 1947; Eady, 1949). However, the lifecycle of hazardous extreme weather is rarely determined by linear growth alone; it is dictated by the nonlinear saturation phase, characterized by wave breaking, vortex interactions, and decay. The theoretical connection between jet latitude and upper-level Rossby Wave Breaking (RWB) paradigms—specifically the transition between anticyclonic (LC1) and cyclonic (LC2) lifecycles—has been documented in idealized frameworks (Thorncroft et al., 1993; Rivière, 2009; Barnes & Hartmann, 2012). The RWB occurs at the mature life stage of baroclinic waves when the upper-level flow's potential vorticity contours overturn, causing potent, irreversible mixing of air masses across latitudes. Since the upper-level dynamics affect the motion of surface weather systems and steers the nonlinear interactions, this nonlinear phase is directly responsible for many high-impact weather events (Martius & Rivière, 2016; Tamarin-Brodsky & Harnik, 2024, Tamarin-Brodsky et al., 2026). For example, cyclonic RWB is often associated with explosive cyclones, leading to extreme low-level wind speeds (Sanders & Gyakum, 1980; Gómara et al., 2014). Anticyclonic RWB is associated with the formation of persistent surface high-pressure systems, causing prolonged heatwaves or cold spells (Kautz et al., 2022). Overall, the sensitivity of these surface weather systems, especially their nonlinear interactions, to the jet structure variations beyond latitudinal displacement received little attention.

One specific class of nonlinear interaction that remains understudied is the cyclone merging. Cyclone merging represents a distinct class of nonlinear interaction that contributes to the "heavy tail" of the wind speed distribution in the observation (Freedman, 2020; Chen et al., 2022) and climate simulations (van den Brink et al., 2004). The coalescence of surface cyclones can lead to explosive intensification, resulting in deeper minimum pressures and stronger wind fields than single-cyclone events (Freedman, 2020). In fact, statistical analyses of extreme synoptic-scale winds in climate simulations reveal a "double population" of wind speeds, where the most violent tail of the distribution is often governed by these complex dynamical interactions rather than standard baroclinic growth (van den Brink et al., 2004). While merging events are associated with some of the most destructive midlatitude storms (Cordeira & Bosart, 2010), the large-scale dynamical conditions that favor their occurrence are difficult to isolate in complex, full-physics model simulations. Existing idealized studies of vortex merging usually focus on idealized two-dimensional vortices (D. G. Dritschel & Waugh, 1992) or vortices in quasi-geostrophic flow without jet streams (David G. Dritschel, 2002; Reinaud & Dritschel, 2018). It remains an open question whether broad jets provide a "low-strain sanctuary" that facilitates these merging events, or if sharp, narrow jets are required to confine the vortices. Establishing a mechanistic link between jet geometry and merging frequency is valuable for interpreting how surface extremes might change in response to structural changes in the large-scale circulation.

On the other hand, the behavior of surface anticyclones represents a critical component of surface weather extremes. The surface anticyclones are dynamically coupled to, yet distinct from, upper-level anticyclones (Zhang et al., 2017; Tamarin-Brodsky & Harnik, 2024). While upper-level anticyclones have been extensively studied, the climatology and variability of synoptic-scale surface high-pressure systems are less well investigated (Pepler et al., 2019), despite their direct control on heat waves, air quality stagnation, and cold air outbreaks (Ioannidou & Yau, 2008). Observational analyses reveal that strong surface anticyclones, such as those originating in Siberia or Alaska, are often driven by radiative cooling and distinct ridge-trough configurations that favor large-scale subsidence (Jones & Cohen, 2011). But to the best of our knowledge, no modeling studies examined how changes in jet structure influence the residence time and intensity of these surface anomalies. Quantifying this sensitivity is essential for determining the relative importance of diabatic processes in the development of surface anticyclones.

Idealized modeling experiments provide valuable tools for isolating and understanding the fundamental dynamics of baroclinic wave evolution, stripping away the complexity of topography and diabatic processes. Historically, these setups have significantly advanced our understanding of baroclinic wave life cycle (e.g., Thorncroft et al. 1993) and helped benchmark dynamical cores (Jablonowski & Williamson, 2006; Ullrich et al., 2017). However, they generally cover only a limited parameter space, often employing just a single or several jet configurations or Cartesian channel domains. The latter, while computationally efficient, imposes artificial zonal periodicity that can constrain the spatial scale of nonlinear interactions, making them less suitable for studying cyclone merging or stationary anticyclones. Recent efforts have sought to overcome these geometric limitations; for instance, Bouvier et al. (2024) introduced an analytical, global-scale initial condition generator for OpenIFS. Meanwhile, the recent containerization of the GFDL FV3 dynamic core (Cheng et al., 2022) provides a portable, scalable framework for global idealized modeling. These technological advancements stage a unique opportunity to move beyond limited case studies and systematically explore how midlatitude jet streams regulate nonlinear interactions across a much broader, physically consistent parameter space.

This study employs the FV3 dynamical core—the non-hydrostatic engine of modern operational forecasting systems—in an idealized dry setting. This approach allows us to bridge the gap between theoretical fluid dynamics and operational modeling. This study focuses on variations in jet latitude, vertical structure, and meridional width and aims to explore the sensitivities of (1) baroclinic wave lifecycle, (2) cyclone merging, and (3) persistent stationary anticyclones. The remainder of this paper is organized as follows. Section 2 describes the model and methodology, such as feature tracking tools. Section 3 presents the simulations of wave evolution and surface weather systems. Section 4 provides a discussion of the broader implications of the findings.

**2 Data and Methods**

This section describes the workflow for generating balanced but baroclinically unstable initial conditions and the configuration of the FV3 dynamical core used to simulate dry baroclinic wave evolution. The modeling framework consists of a three-dimensional (3D) atmospheric model on the sphere, and an adaptable jet stream initial state defined via analytic formulas. We outline: (2.1) the FV3 dynamical core, (2.2) the Python-based initial condition generator and its analytical formulation, (2.3) the setup of perturbations used to trigger baroclinic wave growth, (2.4) the cyclone/anticyclone tracking package ConTrack, and Rossby Wave Breaking identifier WaveBreaking, and (2.5) brief introduction of Eddy Kinetic Energy (EKE).

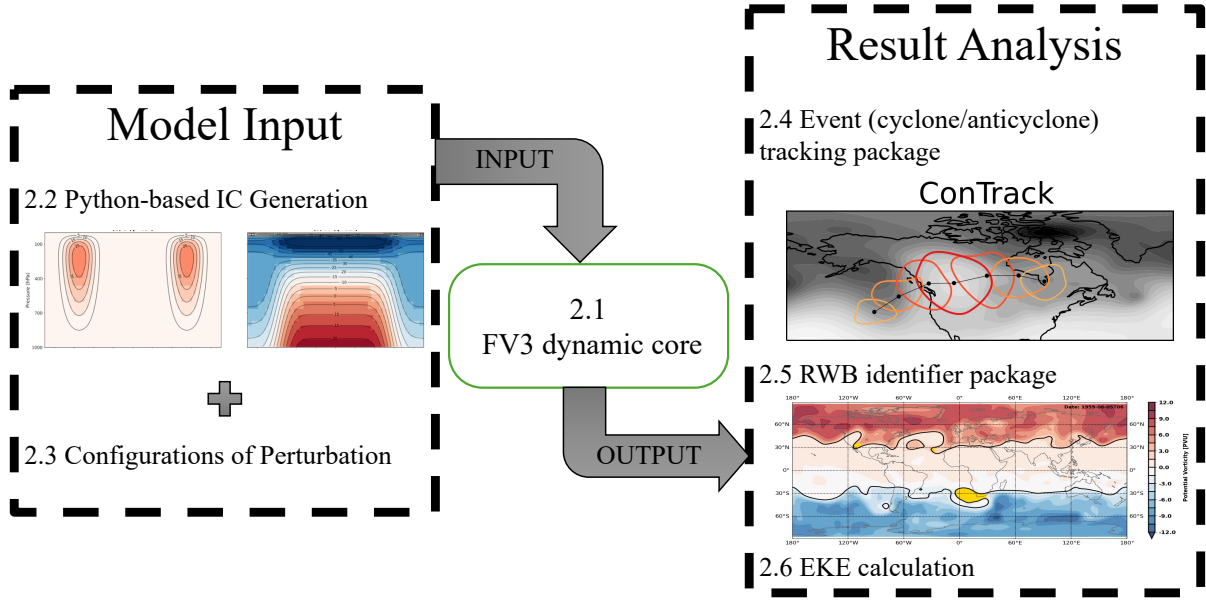

Figure 1. The overall workflow schematic of data generation and analysis.

2.1 Implementation in the FV3 Dynamical Core

We utilize the well-tested GFDL Finite-Volume Cubed-Sphere (FV3) core (Putman & Lin, 2007; Harris & Lin, 2013) to integrate initial conditions forward in time. FV3 integrates the compressible (and, in non-hydrostatic mode, fully compressible non-hydrostatic) Euler equations on a gnomonic cubed-sphere grid, which eliminates polar singularities and provides quasi-uniform spatial resolution over the globe.

For all the experiments, the model is configured as a dry, non-hydrostatic dynamical core without moist physical parameterization and radiation, such as moist convection and precipitation microphysics. This "no-physics" setup serves as a rigorous control, ensuring that any resulting cyclone merging or blocking persistence arise solely from the nonlinear interaction of the background flow and the baroclinic waves, rather than from diabatic feedbacks.

Our experiments adopt a C96 cube-sphere grid of 96×96 cells per cube face, yielding a nominal horizontal grid spacing of approximately ~1.0° or ~100 km at the equator. This resolution is sufficient to resolve synoptic-scale baroclinic instability and large-scale vortex interactions while remaining computationally efficient for the large parameter sweep conducted here. The vertical domain is discretized into 31 levels in a hybrid pressure-sigma coordinate, extending from the surface (1000 hPa) to a constant-pressure (flexible) lid at the top of the atmosphere.

For time integration, we chose a time step of 300 seconds and integrated each experiment for 15 days. The simulation length is generally sufficient to fully capture the linear growth, nonlinear saturation, and decay phases of the baroclinic life cycle. Model output was saved at 15-minute intervals to capture the cyclone evolution and provide energy cycle analysis. We output fields including sea level pressure (SLP), winds, temperature, and geopotential on all levels. Finally, to ensure scientific reproducibility and computational portability, all simulations were executed using the SHiELD-in-a-Box containerized software environment (Cheng et al., 2022). The containerized framework that packages the System for High-resolution prediction on Earth-to-Local Domains (SHiELD) model and its dependencies for reproducible, portable simulations.

2.2 Analytic Jet Initialization and Experimental Design

To investigate the sensitivity of baroclinic wave evolution to background flow geometry, we initialize the model with a suite of analytically specified, zonally symmetric jet streams. Following the formulation of Bouvier et al., 2024, the zonal wind $u(\phi, \eta)$ (where $\phi$ is latitude) is prescribed by an analytic function that allows control over the jet's latitude, width, and vertical structure. In particular, we define the zonal wind profile as:

$$u(\lambda, \phi, \eta) = -u_0 \ln(\eta) \exp\left[-\left(\frac{\ln\eta}{b}\right)^2\right] \sin^{2n}(\phi - s),$$

where $u_0$ controls the maximum of the zonal mean wind speed (set to 35 m/s in the simulations) (Figure 2a-c), $s$ is the latitudinally shifted degree of the jet relative to 45°N, $n$ determines the meridional width of the jet, and $b$ sets the vertical profile of the jet. A larger $n$ yield a sharper, narrower jet centered at $45° + s$. In our formulation, the jet reaches its maximum wind speed at the level $\eta = \exp\frac{\sqrt{2}}{2}b$; increasing $b$ will raise this maximum wind level and yield a vertically deeper jet. This parametric jet profile is highly flexible and can emulate various possible midlatitude zonal wind distributions (Appendix A).

After specifying the zonal wind field $u(\phi, \eta)$, we adjust the geopotential and temperature fields to ensure an approximate balance of forces. In practice, we ensure that the total geopotential field $\Phi(\phi, \eta)$ satisfies the gradient-wind balance equation:

$$\Phi(\lambda, \phi, \eta) = \langle\Phi\rangle(\eta) + \Phi'(\lambda, \phi, \eta)$$

$$\langle\Phi\rangle = \frac{T_{v,0}g}{\gamma}(1 - \eta^{\frac{R_d\gamma}{g}})$$

$$\frac{1}{a}\frac{\partial\Phi'}{\partial\phi} = -u(2\Omega \sin\phi + \frac{u}{a}\tan\phi)$$

where $T_{v,0}$ is the surface virtual temperature, $\gamma$ is the lapse rate, $a$ is Earth's radius and $\Omega$ is the rotation rate of Earth. $\langle\Phi\rangle$ denotes the mean geopotential field on a given vertical level, while $\Phi'$ represents the additional geopotential component required to maintain balance with the zonal wind. This is the zonal momentum equilibrium condition on a rotating sphere (essentially the gradient balance including centrifugal term $\frac{u^2}{a}\tan\phi$). We solve this equation for the geopotential $\Phi(\phi, \eta)$ given the $u(\phi, \eta)$ profile, using an iterative numerical method. We also include a correction to ensure the geostrophic meridional wind is zero initially (Appendix B):

$$\frac{1}{4\pi}\int_0^{2\pi}\int_{-\frac{\pi}{2}}^{\frac{\pi}{2}}\Phi'(\lambda,\phi,\eta)\cos\phi\,d\phi d\lambda = 0$$

The temperature field $T(\phi,\eta)$ is then diagnosed from the hydrostatic relation using the resulting $\Phi$.

$$T_v(\lambda,\phi,\eta) = \langle T_v \rangle + T'_v(\lambda,\phi,\eta)$$

$$\langle T_v \rangle = T_{v,0}\,\eta^{\frac{R_d\Gamma}{g}}$$

$$T'_v(\lambda,\phi,\eta) = -\frac{\eta}{R_d}\frac{\partial\Phi'(\lambda,\phi,\eta)}{\partial\eta}$$

Here $\eta$ is the model's vertical coordinate, $R_d$, the specific gas constant for dry air, and $g$ the gravity constant. The reference surface temperature $T_{v,0}$ and lapse rate $\Gamma$ define the vertical temperature profile in the resting, zonal-mean state.

We verified that our generated base state is steady when integrated forward without additional perturbations (Appendix C). The framework is implemented in Python and allows flexible initialization. We systematically explored a matrix of simulations spanning multiple jet structures by varying the vertical parameter $b$ among {1.0, 1.5, 2.0}, the horizontal width parameter $n$ among {1, 3, 6}, and the jet shift $s$ among {-10°, -5°, 0°, +5°, +10°} (Table 1). Sample flow configurations are available in Figure 2. Our idealized jet configurations were inspired by the global zonal means in the ERA5 reanalysis data (Hersbach et al., 2020). For example, ERA5 indicates the zonal mean midlatitude jet stream in the northern hemisphere shifts between 45°N and 55°N. This design allows us to isolate the marginal effect of jet geometry on extreme statistics, effectively creating a "lookup table" for dynamical sensitivity.

Table 1: Test configurations, including different choices of b (controls the height of the westerly jet), n (controls the width of the westerly jet) and s (controls the shifted degrees of the westerly jet). In total, this yields 3 × 3 × 5 = 45 distinct cases. We use a notation (bX, nY, sZ) to denote a simulation with parameter values b=X n=Y, and jet shift s=Z (in degrees). For example, the "default" case (moderate jet width and unshifted) is (b=2, n=3, s=0), whereas an example extreme case might be (b=2, n=1, s=+10) for a wide jet shifted poleward by 10° at the same height level.

| Parameters | b | n | s |
|---|---|---|---|
| Impacts | Larger values mean jet concentrates at higher altitude (height) | larger values mean jet expands broader meridionally (width) | Larger values mean jet shifts more poleward (latitude) |
| Values | 1.0, 1.5, 2.0 | 1, 3, 6 | -10°, -5°, 0°, +5°, +10° |
| Cardinality | 3 | 3 | 5 |

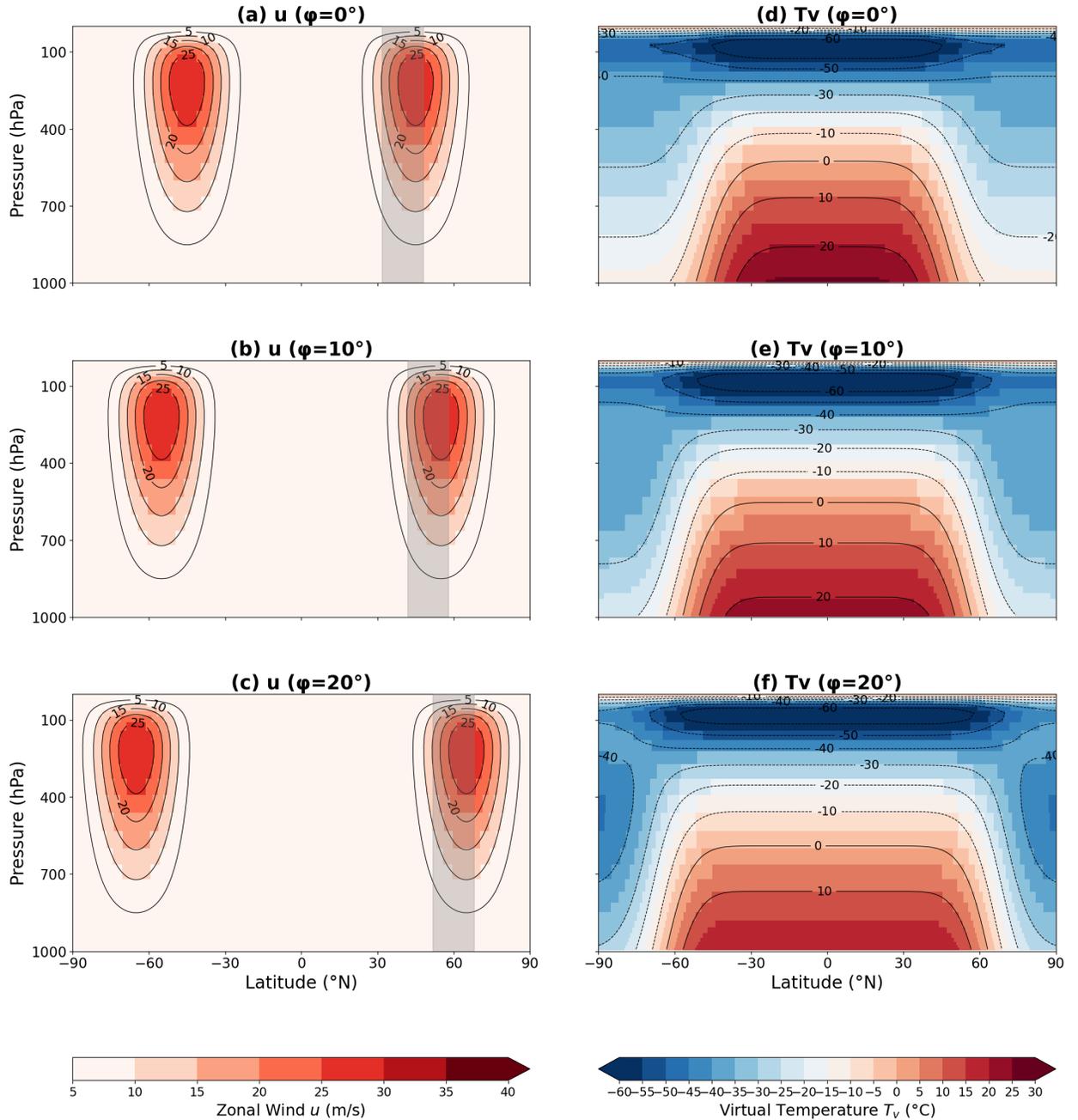

Figure 2: Idealized jet configurations and thermal-wind structure. Left: Zonal wind (u) profiles for (a) a standard midlatitude jet (s=0°), (b) a poleward-shifted jet (s=10°), and (c) a poleward-shifted jet (s=20°). The gray shaded region represents the ±2σ latitudinal extent of the Gaussian-like perturbation. Right: Corresponding cross sections of the virtual temperature showing how the poleward jet modifies the baroclinic zone.

2.3 Perturbation Setup for Baroclinic Wave Initiation

Having established a balanced base state, we introduce a small perturbation to trigger the baroclinic instability and initiate the growth of baroclinic waves. We followed the perturbation strategy by Jablonowski & Williamson (2006) and introduced a localized, Gaussian-shaped perturbation in the low-level zonal wind near the jet core. Specifically, we add a perturbation to the zonal wind of the form:

$$u_\epsilon = u_p \exp\left[-\left(\frac{r}{R}\right)^2\right],$$

where $R = \frac{a}{10}$, $u_p = 1.0 m/s$, r is the great circle distance from the point to the center of the perturbation, given by

$$r = a \arccos(\sin\phi_c \sin\phi + \cos\phi_c \cos\phi \cos(\lambda - \lambda_c)),$$

where $\lambda$ is longitude. For the control case, we center the perturbation at latitude $\phi_c = 40°N$, longitude $\lambda_c = 20°E$. In the other cases, the perturbation was placed 5° latitude south of the jet core in each case (to perturb the baroclinic zone). We move the perturbation center $\phi_c$ along with the jet, keeping the perturbation's position relative to the jet core constant. We conducted additional sensitivity experiments and confirmed that fixing the geographic location of perturbations as the control case yields similar results. An additional sensitivity test with the perturbation placed north of the jet shows that the overall dynamical relationships remain unchanged—specifically, the dependence of RWB frequency on jet structure (see Section 3.1) is preserved. However, the total number of RWB events decreases substantially when the perturbation resides north of the jet stream.

2.4 Feature Identification with WaveBreaking and ConTrack Packages

We adopt the WaveBreaking package to identify and analyze Rossby wave breaking (RWB). The package tracks three characteristic features: PV streamers, overturnings, and cutoffs (Kaderli, 2025). In this study, we use WaveBreaking to diagnose RWB events from potential vorticity (PV) fields. We focus on overturning-type wave breaking by tracking the θ = 330 K contour on the PV = 2 PVU surface. An RWB event is identified when this PV contour overturns meridionally, crossing the same meridian three times. The package captures individual RWB events and records their geometric properties (e.g., area, location) as well as their dynamical type, distinguishing cyclonic from anticyclonic wave breaking.

Surface cyclones and anticyclones were objectively identified and tracked using the ConTrack package (Steinfeld, 2020), an open-source contour-based feature-tracking framework designed for analyzing geophysical fields such as mean sea-level pressure or geopotential height. ConTrack detects closed contours at each time step, links them temporally based on spatial overlap, and assigns a unique identifier to each coherent system through its lifecycle. Three key parameters control the tracking: (1) the variable (pressure in this study) threshold defining candidate contour features, (2) the minimum spatial overlap ratio required between consecutive time steps, and (3) the persistence criterion specifying the minimum duration for a valid track.

To isolate cyclone merging and persistent anticyclones, we applied ConTrack to hourly surface pressure fields to identify and classify transient and stationary systems. Cyclone merging events were defined when two independent 990 hPa contours merged into a single coherent feature. Stationary anticyclones, in contrast, were identified from the 1010 hPa isobar, requiring at least 0.8 spatial overlap between successive time steps and a minimum persistence of 72 h.

2.5 Eddy Kinetic Energy

Eddy kinetic energy (EKE) is a widely used measurement of the intensity of transient, synoptic-scale atmospheric motions to evaluate the activity level of weather systems like midlatitude cyclones and their associated eddies. It is defined as the wind anomalies relative to the mean flow,

$$EKE = \frac{1}{2}(u'^2 + v'^2),$$

where $u'$ and $v'$ are zonal and meridional wind perturbations.

**3 Results**

3.1 Baroclinic Wave Lifecycle and Energy Evolution

Here, we examine the lifecycle of developing baroclinic waves, the evolution of EKE, and the statistical distribution of RWB events under varying initial jet configurations. Our analysis starts with a baseline case with a jet centered at 45°N (b=2, n=3, s=0) and a case where the initial jet is displaced poleward by 10° to 55°N (b=2, n=3, s=+10).

The early-stage evolution of the surface pressure and 850-hPa temperature fields reveals the classic signatures of baroclinic instability (Figure 3). By Day 8 of the baseline simulation (Figure 3a-b), a train of surface cyclones (low-pressure centers) and anticyclones (high-pressure centers) has emerged. These systems exhibit the characteristic thermal structure of developing waves, with strong warm air advection ahead of each surface low and cold air advection in its wake, consistent with the standard lifecycles documented in idealized studies and in observed synoptic systems (e.g., Thorcroft et al., 1993; Polvani & Esler, 2007).

Comparing the baseline and polar-shifted jets cases reveals notable differences in the intensification and movement of the primary cyclone (Figure 3f and 3g). In the polar-shifted case, the cyclones deepen more rapidly and reach lower central pressures than their counterparts in the baseline experiment. The warm and cold fronts around each cyclone are also stronger (Figure 3g). The polar-shifted jet yields faster and stronger cyclogenesis: by this time (Day 6), the primary cyclone (noted as P) in Figure 3f is more intense and has advected farther eastward compared to the corresponding cyclone in the baseline case (Figure 3a).

As the simulations progress into the nonlinear regime, the phase speed differences become distinct and create an environment conducive to vortex interactions. For example, the primary cyclone **P** is observed to merge with a trailing cyclone between Day 10 (Figure 3c) and Day 12 (Figure 3e). The cyclones propagate northeastward relative to the anticyclones (Figure 3g-j). Compared with cyclones, anticyclones move more slowly, covering fewer degrees of longitude over the same simulation period; in some cases, they remain nearly stationary. The cyclone interactions and stationary anticyclones will be further discussed in Section 3b and Section 3c.

The development of an upper-level wave pattern provides insights into the baroclinic nature of these disturbances (Figure 4). In all cases, deep upper-level troughs (marked by concentrated positive vorticity) align to the west of surface lows, indicating the westward tilt with height required for baroclinic growth. However, the geometry of the jet strongly modulates the wave breaking morphology. In the baseline/equatorward configurations (e.g., Figure 4d), the upper-level troughs exhibit a pronounced northwest–southeast tilt, indicative of cyclonic wave breaking (CWB). Conversely, the poleward-shifted case (Figure 4h) exhibits troughs and ridges with a northeast–southwest tilt, a signature of anticyclonic wave breaking (AWB). This structural dichotomy suggests that equatorward jets favor cyclonic overturning (LC2), while poleward jets transition the system toward anticyclonic (LC1) behavior (Thorncroft et al 1993; Rivière, 2009).

To quantify this sensitivity, Figure 5 presents the statistical frequency of RWB types across the parameter sweep. While Anticyclonic RWB is the dominant mode across most configurations, the partition is sensitive to jet geometry. We find that broader jets (increasing n) allow a slightly higher fraction of CWB. Conversely, increasing jet height (larger b) systematically suppresses cyclonic overturning, reinforcing the AWB preference. The most pronounced sensitivity appears in the latitude-shift suite: equatorward-shifted jets (s= -10, -5) favor CWB, while poleward shifts (s= 5, 10) rapidly transition the storm track to an exclusively AWB regime. Across the sweep, the total number of RWB events (numbers annotated within each bar) ranges from 408 to 1947 and varies most strongly in the width (n) and latitude-shift (s) suites (e.g., 1947 at n=1 versus 889 at n=6, and 408 at s=-10 versus 1083 at s=10). Together, these results indicate that jet structure strongly modulates not only the cyclonic–anticyclonic partition but also the total number of RWB events, with poleward and higher jets particularly conducive to anticyclonic RWB (Garfinkel & Waugh, 2014).

Finally, we analyze the evolution of domain-averaged eddy kinetic energy (EKE) to characterize the bulk intensity of the storm tracks (Figure 6). Starting from 192h (Day 8), all cases undergo the development of baroclinic waves, characterized by an initial rapid increase in EKE, which lasts for around 96h (4 days). The influence of jet position is evident in the growth phase. The polar-shifted cases (the s=10 experiment) generally show a steeper rise in EKE initially and attain 13 $m^2/s^2$ on Day 10 (Figure 6c, b2n1s10 case) sooner than their unshifted counterparts. This aligns with our earlier observation of more rapid cyclone deepening in the polar jet configuration. Yuval et al. (2018) reported enhanced baroclinic growth and increased EKE activities under a poleward-shifted jet configuration in both reanalysis and idealized GCM simulations, which is consistent with our finding. Nonetheless, the EKE sensitivity to the jet width and depth can be comparable to or even larger than the sensitivity to jet latitude in the examined parameter space (e.g., Figure 6c vs 6g).

In summary, the experiment results are qualitatively consistent with past studies of the wave evolution dependencies on the background flow (e.g., Simmons & Hoskins, 1978; Thorncroft et al., 1993). Specifically, the simulation results show that a poleward-shifted jet leads to faster initial cyclone growth, earlier peaks in intensity, and an AWB preference. The simulations also reveal new insights. Notably, broader and deeper jets favor more rapid, amplified wave development with more AWB events. Having established this valid dynamical baseline, we proceed to examine how these structural changes regulate the specific nonlinear extremes of cyclone merging and stationary anticyclones.

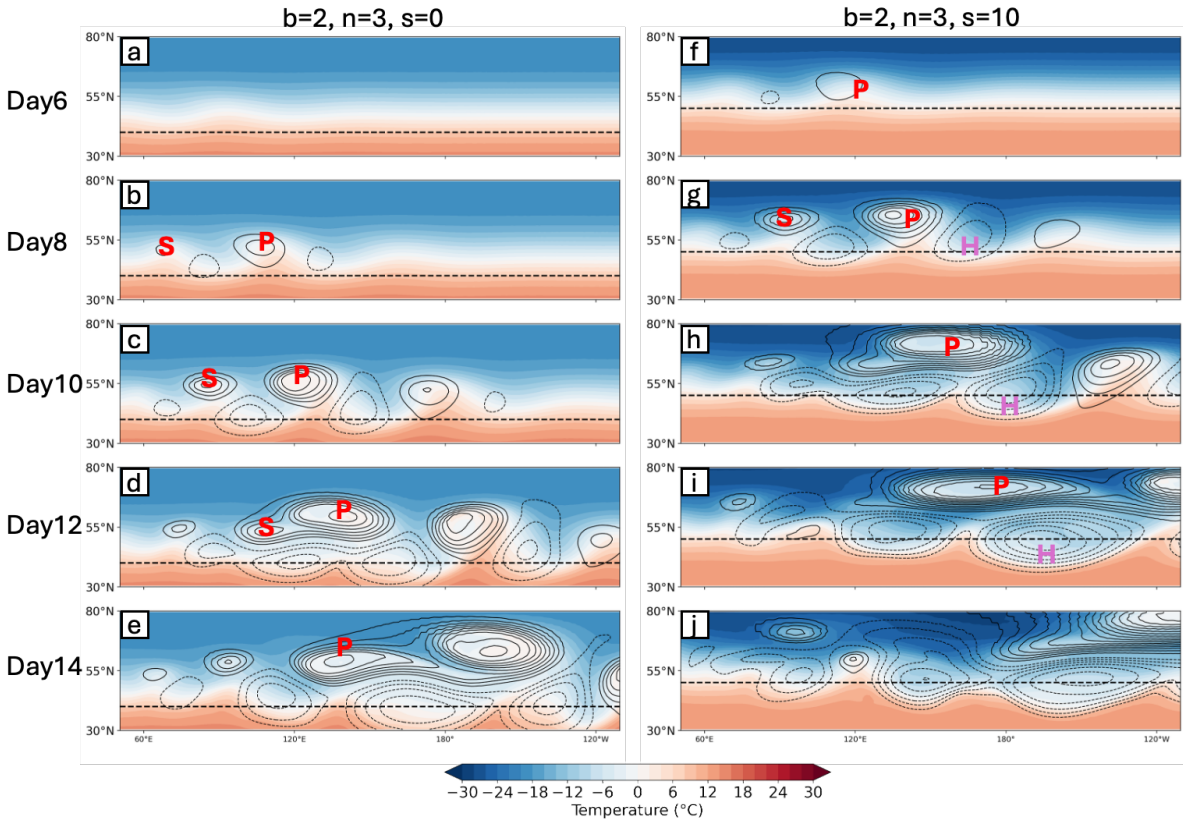

Figure 3. Example cyclone life cycle under idealized jets on even-numbered days from Day 6 to Day 14. For a baroclinic wave evolving, surface pressure (contour) and 850 hPa temperature (shading) are shown. Solid and dashed lines denote isobars for pressures below and above 1000 hPa, with a contour interval of 10 hPa. The dashed black horizontal line indicates the latitude of the initial jet. (a) – (e) for the b2n3s0 case, (f) – (j) for the b2n3s10 case. The primary cyclone was noted as red 'P', and the secondary cyclone was noted as red 'S', while the high mentioned in Section 3a is emphasized as a purple 'H'.

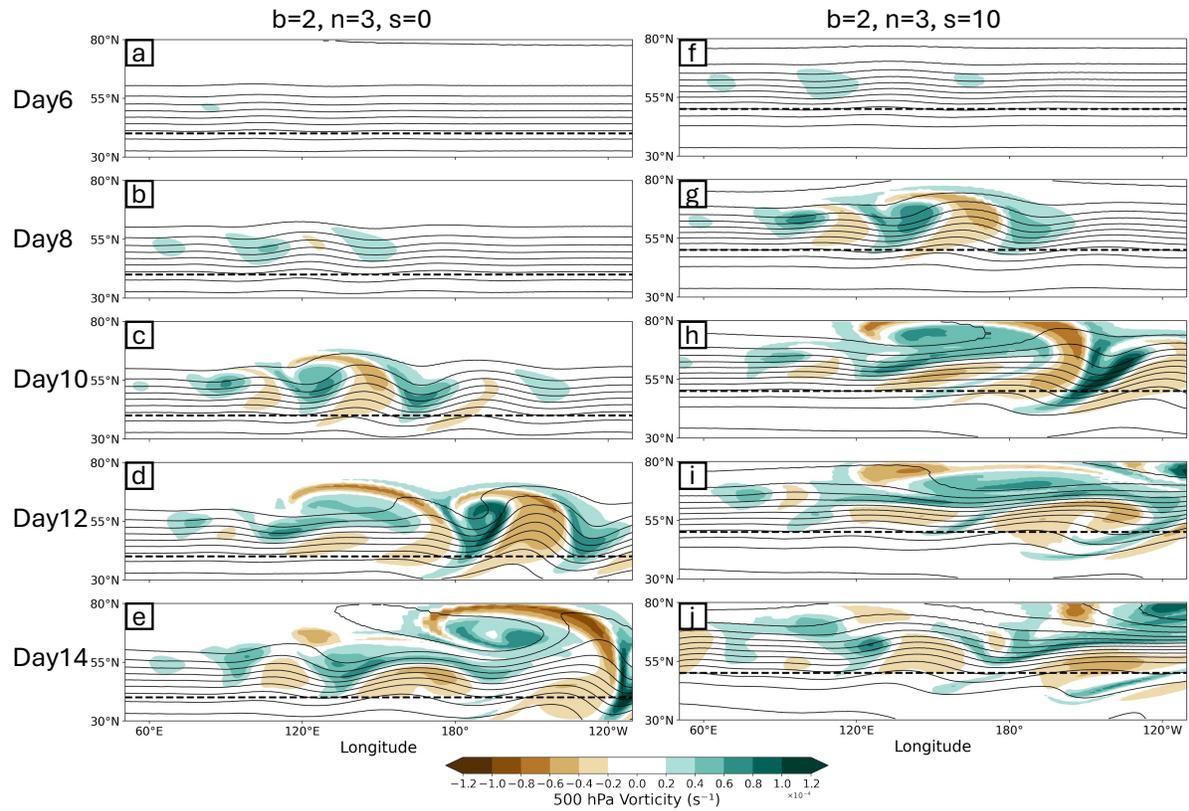

Figure 4. Upper-level flow evolution. Shown are 300-hPa geopotential height fields (contour) and 500-hPa vorticity (shading) for an evolving baroclinic trough under a fixed jet. (a) – (e) for the b2n3s0 case, (f) – (j) for the b2n3s10 case.

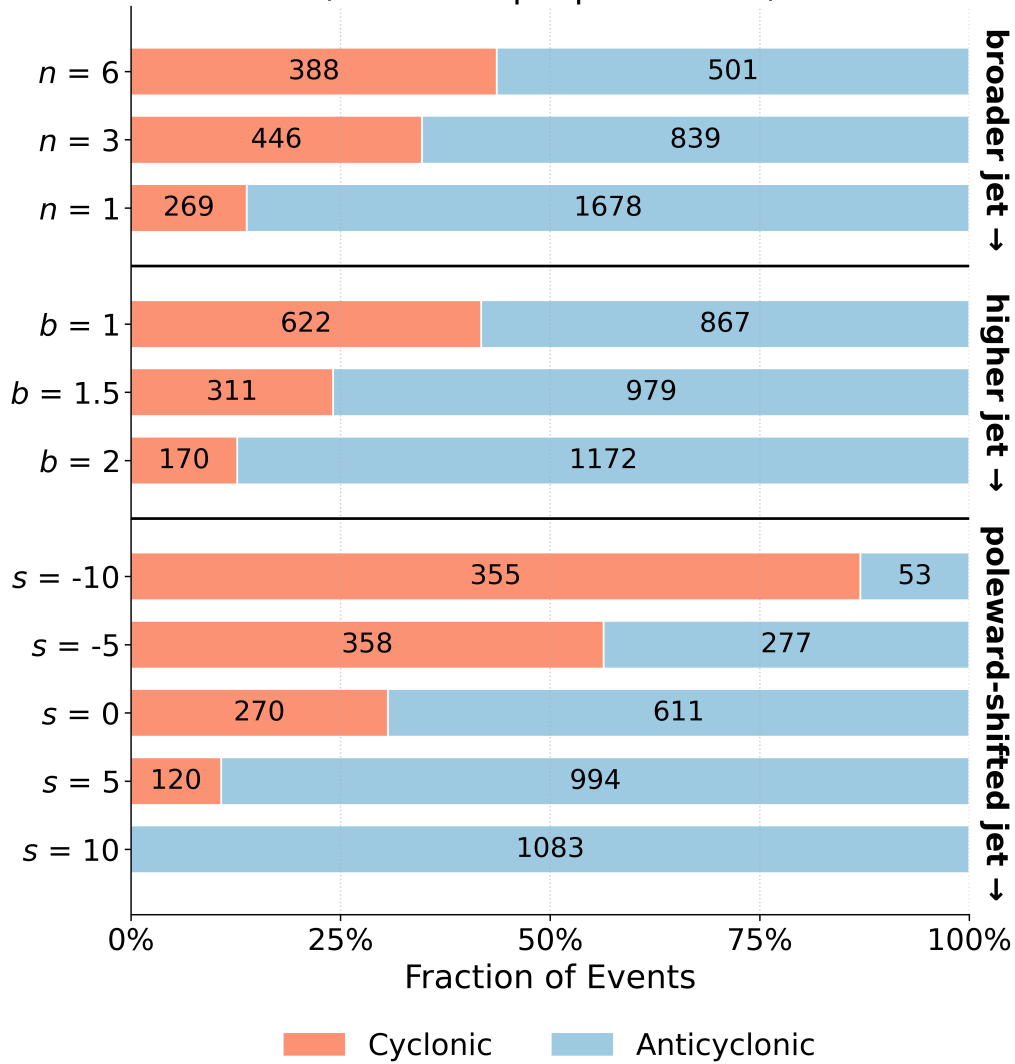

Figure 5. RWB statistics diagnosed by WaveBreaking under different jet configurations from all simulations. There are 4,121 RWB events captured across all 45 cases in the 15-day simulations. From top to bottom, the figure shows experiments with (i) varying jet width parameter n, (ii) varying jet height/depth parameter b, and (iii) varying jet latitude-shift parameter s. Each horizontal bar is stacked to show the fraction of diagnosed overturning cyclonic RWB events versus anticyclonic events, with the numbers inside each segment indicating the corresponding event counts.

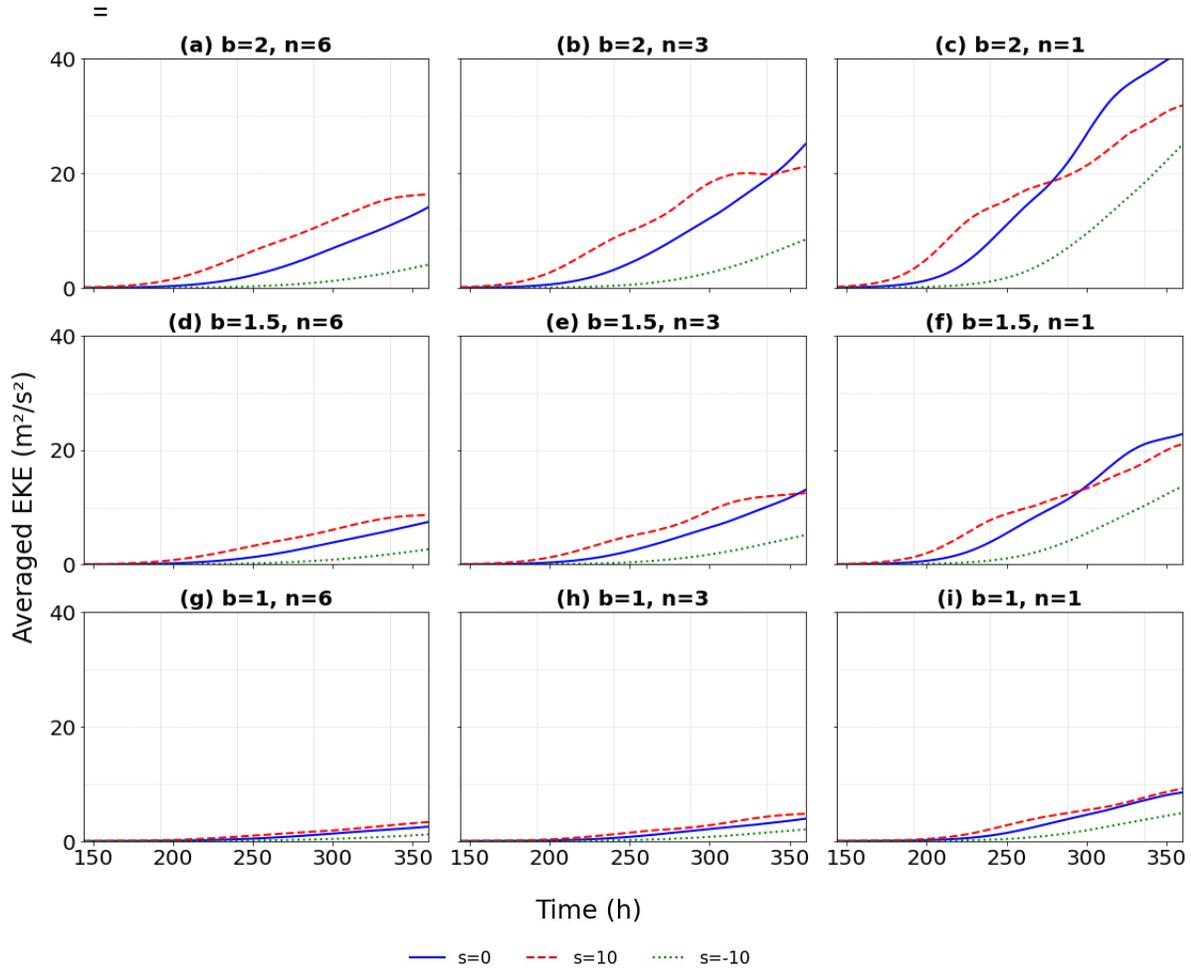

Figure 6. Eddy kinetic energy (EKE) evolution across experiments. Domain-averaged EKE (m^2/s^2) as a function of time for different jet configurations. The curves demonstrate that more poleward-shifted (larger s), broader (smaller n), and higher (larger b) jets tend to produce faster and larger EKE growth, whereas more equatorward-shifted (smaller s), narrower (larger n), and lower (smaller b) jets yield slower, weaker growth.

3.2 Cyclone Merging and Wind Extremes

In our simulations, cyclone merging events are not random anomalies but are systematically regulated by the background jet geometry. The influence of jet structure on the morphology of these interactions is visually evident in simulation cases. In the baseline control case (s=0) (Figures 7a-c), a primary cyclone interacts with a trailing cyclone approaching from the west. The resulting merging occurs between Day 12 (Figure 7b) and Day 14 (Figure 7c), with the trailing cyclone sheared out and eventually absorbed. However, shifting the jet poleward (s=+10; Figures 7d-f) accelerates this timing. The primary cyclone intensifies more rapidly, developing a deeper surface low and stronger vertical motion compared with the baseline case. This creates a steering flow that quickly entrains the weaker trailing system (Figure 7e). The most extreme interactions, however, emerge under the broadest, poleward-shifted jet configurations (b2n1s10 scenario; Figures 7g-i). The primary cyclone develops quickly and evolves into a polar cyclone, sustaining its intensity through repeated absorption of adjacent cyclonic systems. The primary system rapidly entrains a trailing cyclone (Figure 7g) and subsequently captures a second downstream system (Figure 7i), a multi-stage merging process that prevents the typical decay phase observed in narrower jet regimes.

The dynamical consequence of these merging is a sharp, transient spike in surface wind speeds, driven fundamentally by the conservation of angular momentum. As the effective radius of the binary system contracts during coalescence, wind speeds in the interface region surge. Taking the b2n3s10 case as an example (Figure 8a-c), we observe that the "saddle point" between merging vortices becomes a locus for extreme winds. On day 10, the maximum wind speed reached 50 m/s, which is the highest surface wind speed in the first 11-day simulation in this case. A composite analysis of all identified merging events (Figures 8d–f) confirms that this amplification is systematic, with localized wind maxima typically peaking between 0 and +12 hours relative to the merging time. Crucially, while the latitude of the jet has a negligible effect on the absolute magnitude of this wind spike (<5 m s$^{-1}$ difference across shifts, Figure 8d), the vertical and horizontal structure are dominant controls. Deeper jets (increasing b) and broader jets (n=1) consistently produce stronger peak winds than their shallower or narrower counterparts, likely due to reduced baroclinic decay rates in the lower troposphere, which extend the dynamical longevity of the individual cyclones and increase the frequency and intensity of cyclone merging.

Beyond the intensity of individual events, the *frequency* of these interactions is governed by a distinct geometric mechanism, revealing that broad jets facilitate frequent and distinct track coalescence, whereas narrow jets often result in parallel tracks that fail to merge. This behavior suggests a "low strain sanctuary": narrow jets impose strong background shear that stretches and separates adjacent vortices, while broad jets create a kinematic environment of low deformation that allows vortices to rotate around each other and merge. The Hovmöller diagram (Figure 9) provides a qualitative visualization of this regime shift, illustrating that deeper and broader jets foster a higher density of converging tracks. However, it is important to note that the visual convergence of tracks in a Hovmöller diagram does not, by itself, confirm a physical cyclone merging. Algorithmic feature tracking (Figure 10) provides concise quantitative confirmation of this regime shift, showing a pronounced sensitivity of merging frequency to the jet's latitude (s), width (n), and height (b). Summing all the event counts with various jet latitudes, an equatorward jet (s = –10°) produces only 6 merging events across the full integration, whereas a poleward jet (s = 10°) yields 56 merging events. Furthermore, for any fixed latitude, increasing the jet width consistently amplifies the merging count, confirming that broad, poleward jets provide the optimal dynamical conditions for vortex coalescence.

In conclusion, jet geometry acts as a control switch for the merging regime. Poleward-shifted, broader, and higher jets promote stronger baroclinic interactions and repeated vortex coalescence. For the same jet intensity, this synergy suggests that changes in jet structure could shift the distribution of extremes toward the merging-driven tail identified in climate simulations.

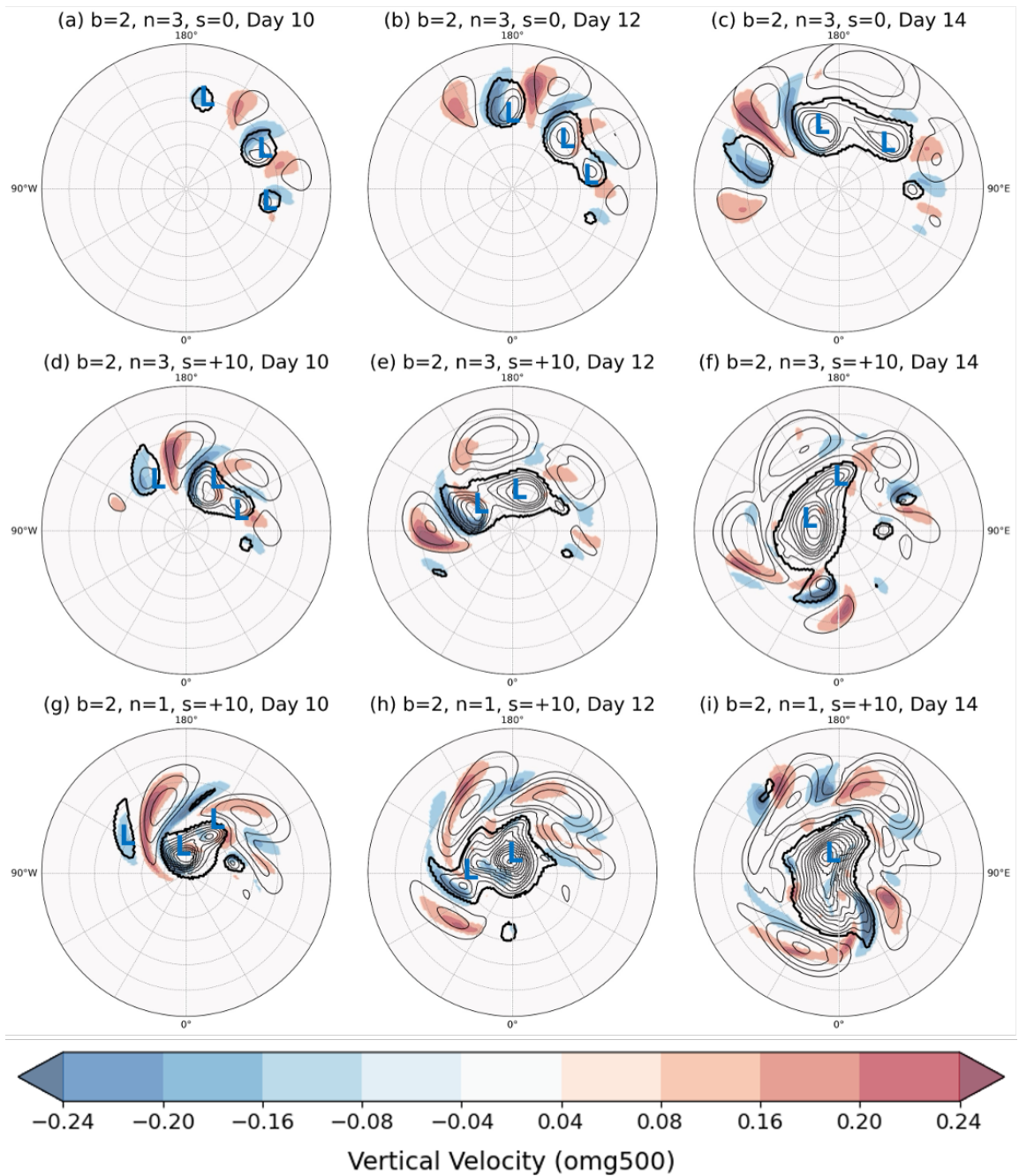

Figure 7. Northern Hemisphere polar stereographic projections depict the 500-hPa vertical velocity (omega) and surface pressure (contours, interval 10 hPa) at days 10, 12, and 14 for three simulation scenarios: b=2 n=3 s=0 (a), (b), and (c), b=2 n=3 s=+10 (d), (e), and (f), and b=2 n=1 s=+10 (g), (h), and (i). The cyclones noted with 'L' are the primary cyclones and their neighborhoods in the downstream and upstream regions. The bold black contour indicates the cyclone tracked with ConTrack. The figures illustrate the progressive merging of cyclones, with the strongest merging dynamics evident in the b2n1s10 case, where a dominant primary cyclone rapidly absorbs adjacent weaker cyclones.

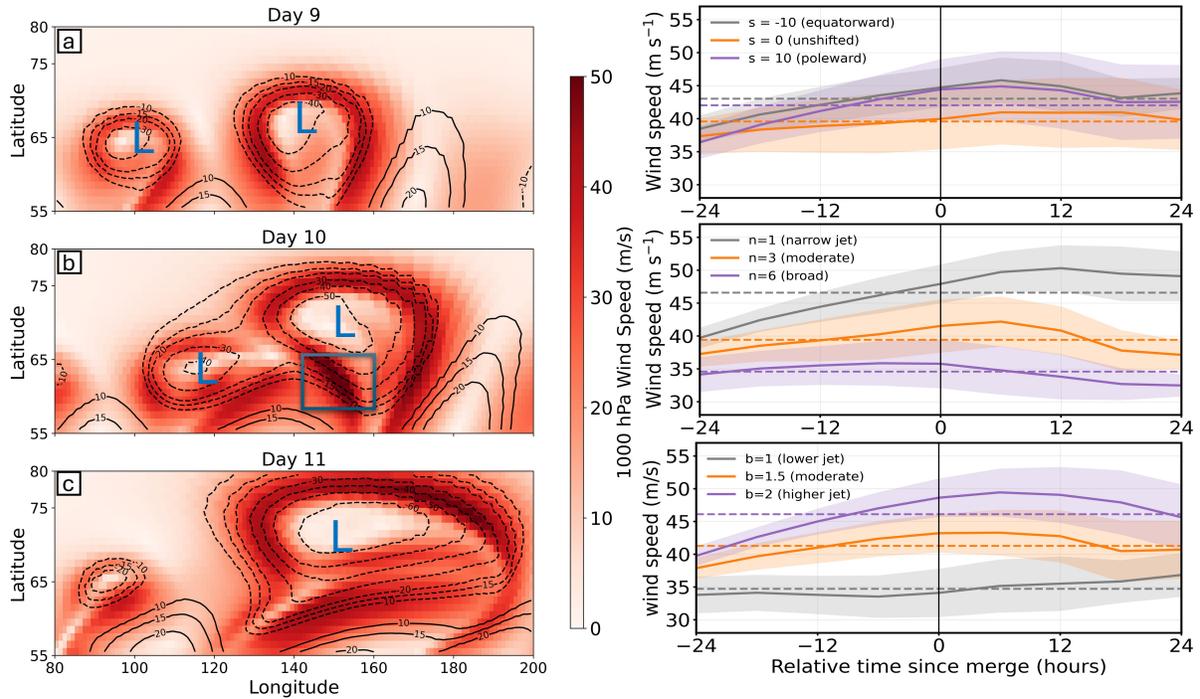

Figure 8. Evolution of surface winds and cyclone merging. (a–c) 1000-hPa wind speed (shading) and surface pressure (contours) for case b=2,n=3,s=10, showing the wind maximum during cyclone merging. (d–f) Maximum 1000-hPa wind speed relative to the merging time (t=0, vertical line), defined as the first time the primary and the trailing cyclone become a single connected system. Sensitivity is shown for (d) jet latitude, (e) jet width n, and (f) vertical concentration b. Shaded envelopes denote ±0.5σ ensemble variability; dashed lines indicate group means.

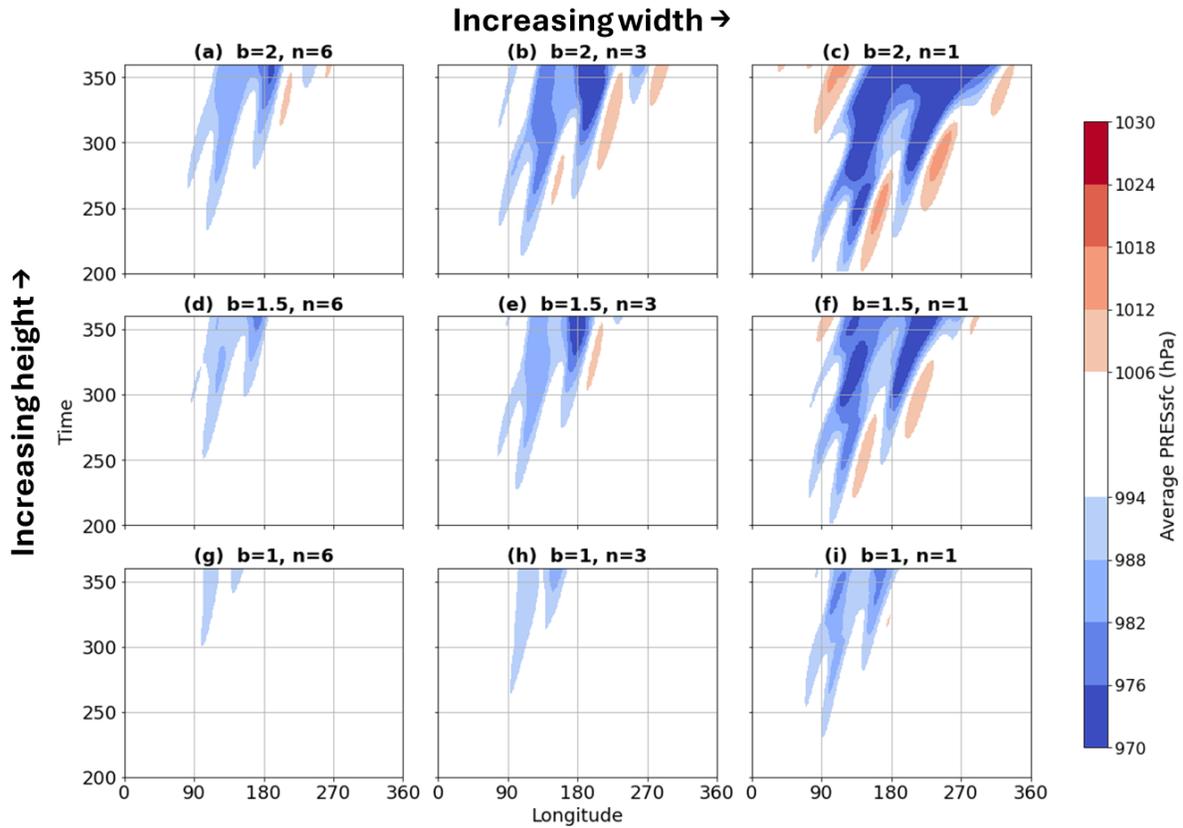

Figure 9. Hovmöller diagram (longitude-time) of latitude-averaged surface pressure (50–90°N) for baseline s=0 scenarios (no poleward shift). In (a)-(c), (e), (f), (i), distinct low-pressure tracks gradually converge longitudinally, indicating the potential merging between the primary low-pressure system and the low-pressure system in the upstream region at different times. In other cases, the phenomenon is not obvious.

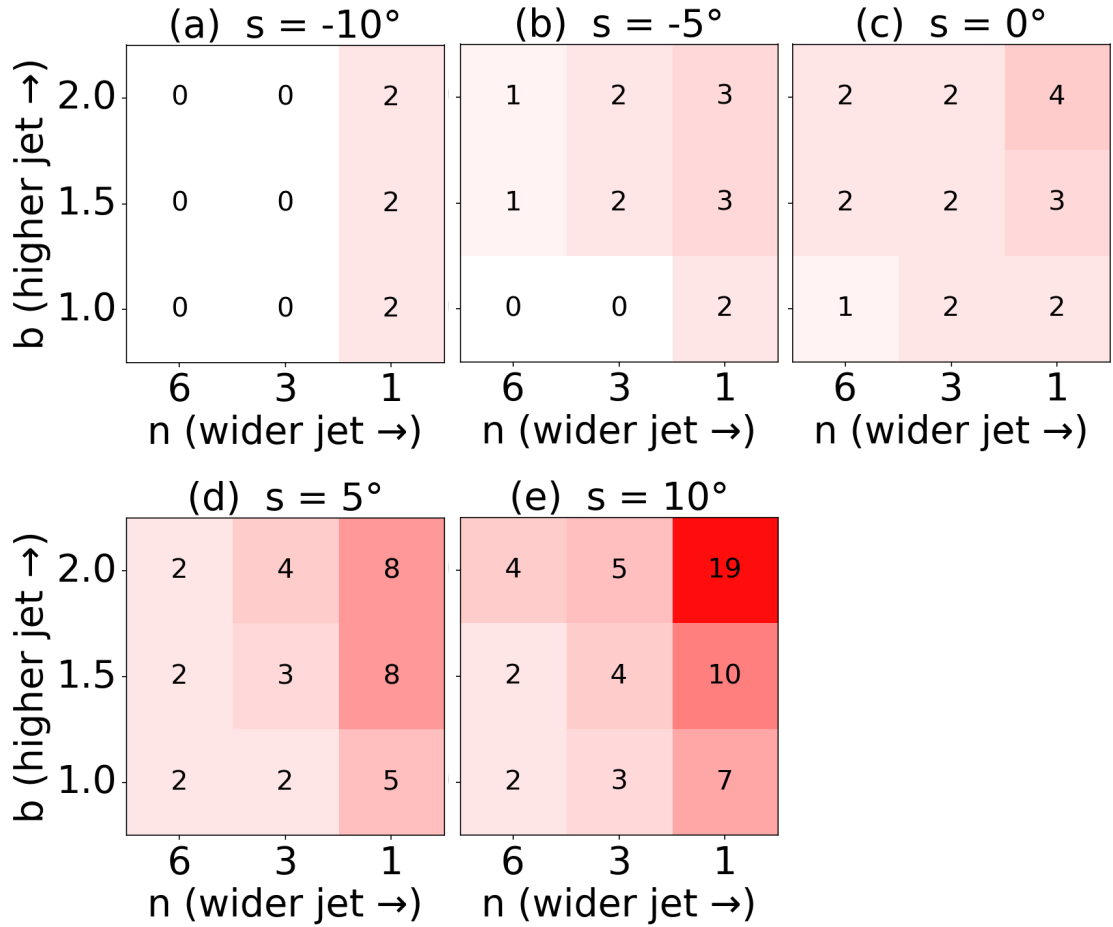

Figure 10. Number of cyclone-merging events as a function of jet height (b; y-axis) and jet width (n; x-axis) for five jet-latitude configurations (s = −10°, −5°, 0°, 5°, 10°) in the 15-day simulations. Each panel shows how varying the jet's meridional position and structure modulates the frequency of cyclone merging in the idealized simulations. For southward-shifted jets (s ≤ 0°; panels a–c), merging activity remains weak and largely confined to narrow, low-level jets. As the jet shifts poleward (s = 5° and 10°; panels d–e), merging occurs more frequently.

3.3 Stationary Anticyclones

In this section, we define a stationary anticyclone as a surface high-pressure system that remains nearly fixed in location for 3 days, often associated with heatwaves and cold spells. In our idealized simulations, stationary anticyclones emerge spontaneously from the background flow dynamics, yet their formation and longevity are strictly conditioned by the initial geometry of the jet stream.

The morphology of these features is illustrated in Figure 11, which tracks the evolution of the deep, broad, and poleward-shifted jet scenario. A large high-pressure center emerges around ~55° N 120° E by day 7 and remains nearly stationary over the same region for the following five days. This anticyclone is evident as a bull's-eye of elevated geopotential height surrounded by closed contours, co-located with a broad area of subsidence characteristic of a high-pressure ridge. To distinguish such stationary features from propagating waves, we applied a Lagrangian tracking algorithm with a strict overlap criterion: a high-pressure center is deemed as stationary only if its closed pressure contours maintain at least 80% areal overlap between consecutive time steps (1h). In the b1n1s10 case, the movement speed of the tracked anticyclone from Day 10 to Day 14 is 3.1 m/s.

Although our dry simulations preclude the thermodynamic feedbacks typically involved in blocking maintenance, the circulation driven by these stationary anticyclones is sufficient to generate significant temperature anomalies through advection. Figure 11g illustrates the zoomed-in view of the b2n1s10 case as an example, where two stationary anticyclones persist for multiple days. Persistent anticyclonic flow creates notable cold anomalies on the southeastern edge of the anticyclone and warm anomalies on the northwestern edge. This evolution demonstrates that the flow geometry itself can precondition the atmosphere for prolonged local heat and cold extremes, independent of moisture-driven amplification.

Figure 12 presents Hovmöller diagrams of surface level pressure (averaged over 0–50° N) for the s=10 simulations. In the b2n1s10 run (Figure 12c), the subtropical anticyclone manifests as an almost vertical streak in the time–longitude plane, indicating that the zonal phase speed of the Rossby wave packet approaches zero. This contrasts sharply with the baseline and equatorward cases, where anticyclones exhibit the flatter slopes characteristic of eastward propagation. A poleward shift of the jet displaces the waveguide into a region where the planetary vorticity gradient $\beta$ is weaker. This reduction in the effective restoring force, together with broader jets support the growth of larger zonal wavelengths (low wavenumbers), which makes the zonal propagation more slowly against the mean westerly flow according to the linear Rossby wave theory (Pelly & Hoskins, 2003). We also speculate that a higher tropopause allows for deeper columns of potential vorticity anomalies, making the surface anomalies harder to displace. These physical mechanisms warrant more investigation.

Figure 13 provides the detailed sensitivity of the primary stationary anticyclone duration across the jet-parameter space (s, n, b). The sensitivity to latitude is particularly pronounced. For equatorward jets (s=-10, 5), several configurations never produce a stationary anticyclone (0 h), indicating strong dependence on jet latitude. With a poleward shift (s=10), stationary events become ubiquitous and sensitive to jet structure. We observe that deeper and broader jets maximize persistence, with the duration of the anticyclone increasing from 121 h in the shallow–narrow case (b1n6s10) to 199 h in the deep–broad case (b2n1s10). This systematic response confirms that naturally occurring or forced expansions of the jet width and depth can purely kinematically precondition the atmosphere for longer-duration blocking events.

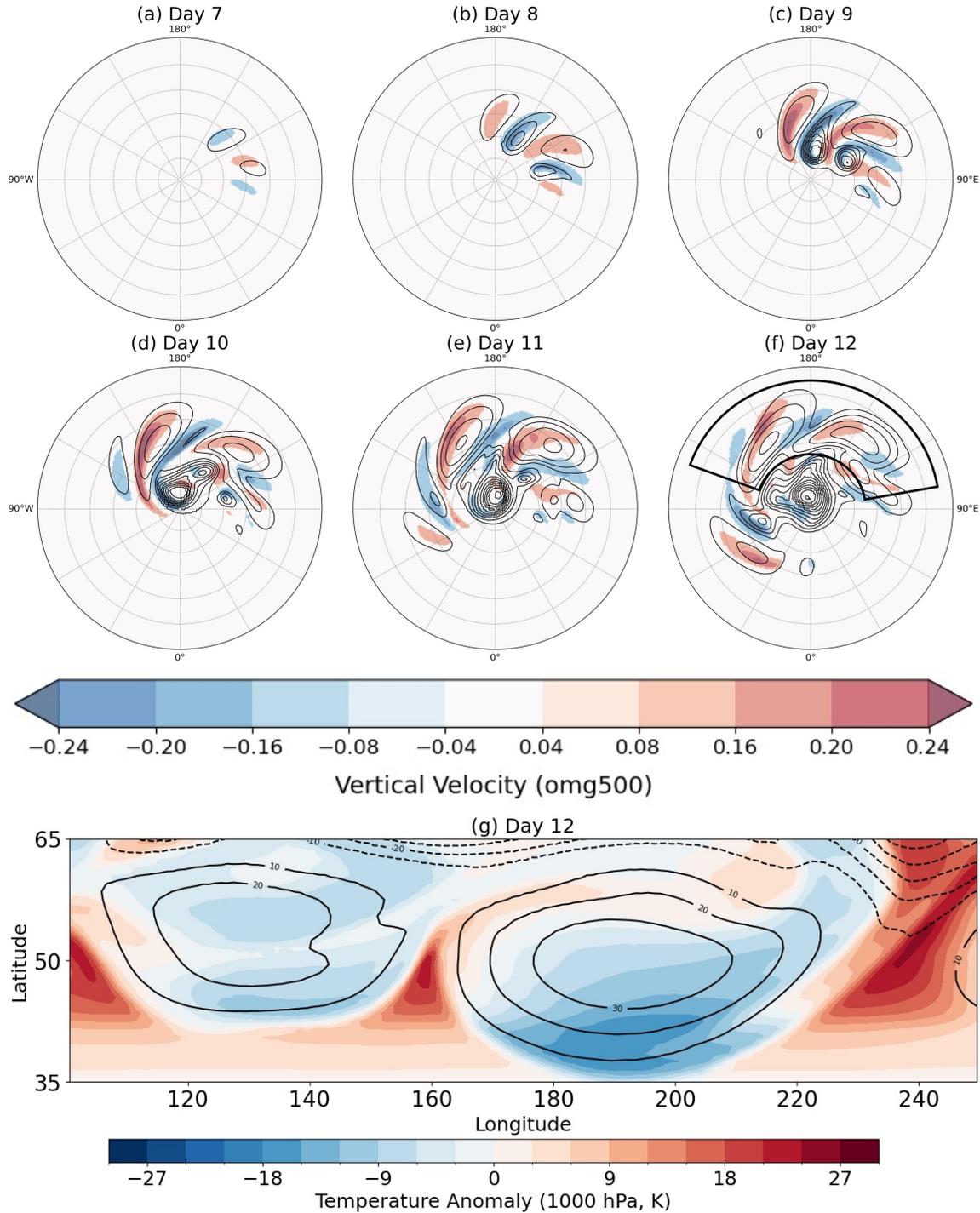

Figure 11: Panels (a-f) present a series of Northern Hemisphere polar stereographic plots displaying the evolution of the surface pressure (contours, interval 10 hPa) and vertical motion (omega) fields over days 7–12 for the simulation scenario with parameters b=2, n=1, s=+10. The persistent, stationary high-pressure center noted as 'H' is clearly visible throughout this period, indicating a robust stationary anticyclone event characterized by minimal longitudinal displacement and pronounced anticyclonic flow. In panels (d) and (f), the thick black box marks

manuscript submitted to *Journal of Advances in Modeling Earth Systems (JAMES)*the zoom-in domain shown in the subsequent panels (g) and (h). Panel (g): Stationary anticyclone in case b2n1s10 with temperature-anomaly field. Solid contours denote surface pressure; shading shows 1000-hPa temperature anomaly (K).

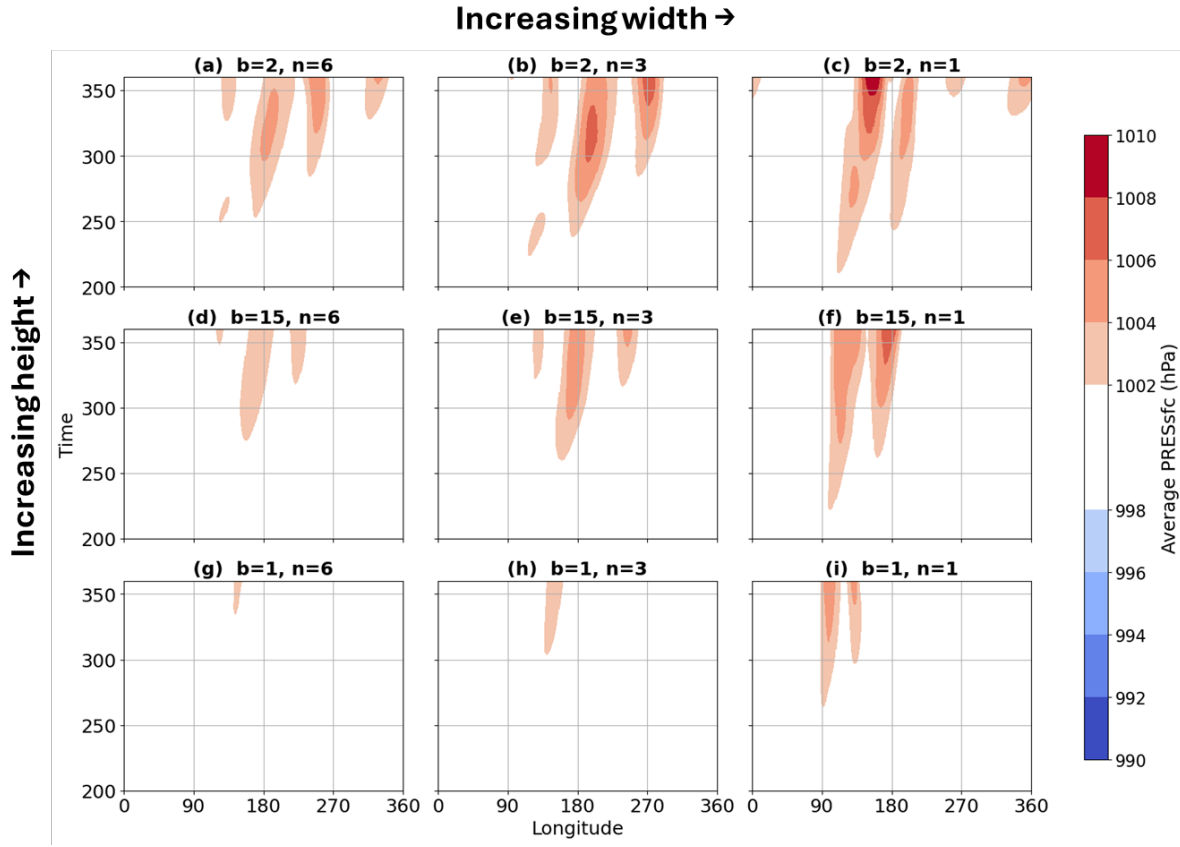

Figure 12. Hovmöller diagrams (longitude-time plots) of latitude-averaged (0–50°N) surface pressure multiple simulation cases, focusing on scenarios with a poleward-shifted jet (s=+10). Notably, the b2n1s10 case demonstrates a distinctly stationary high, depicted as a nearly horizontal streak extending through multiple consecutive days, emphasizing the temporal persistence and stability of the stationary feature.

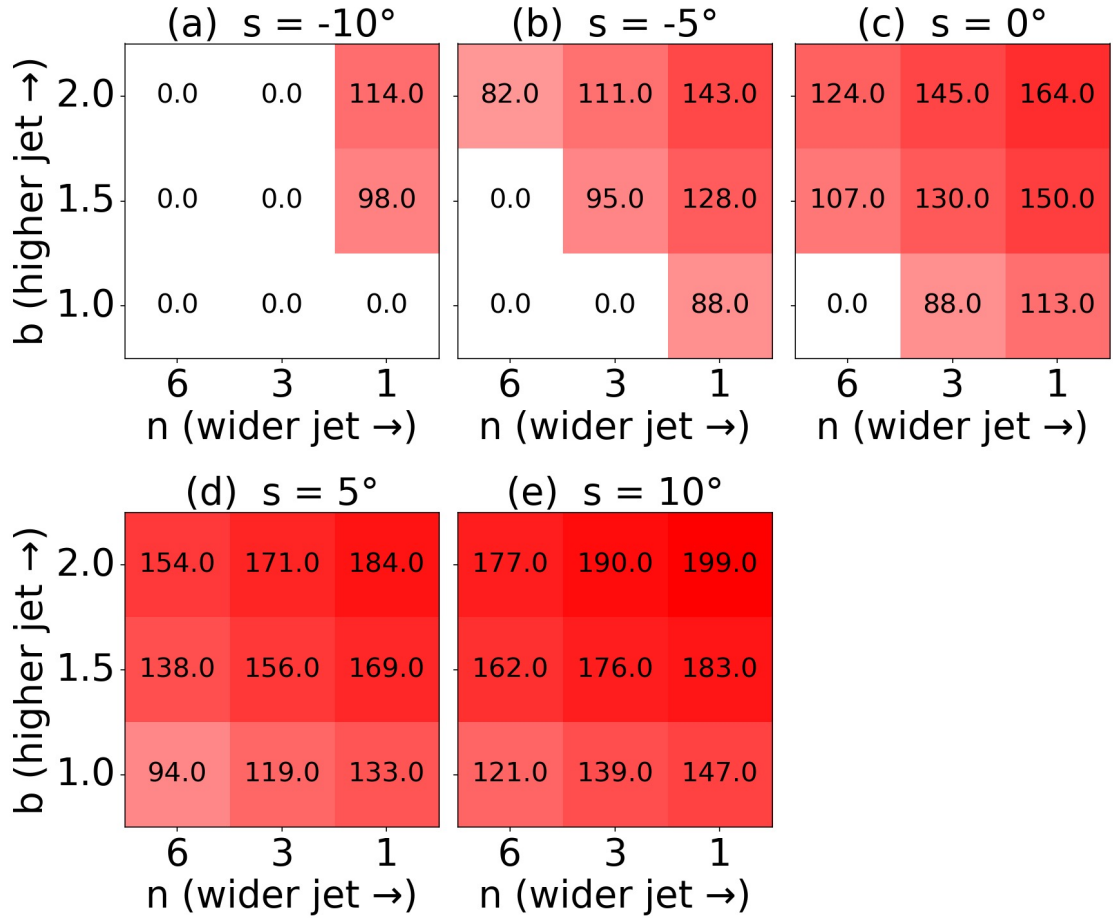

Figure 13. The sensitivity of the primary stationary anticyclone duration to jet configuration parameters using a heatmap. Primary anticyclone is the first stationary anticyclone event identified and tracked by ConTrack based on the prescribed threshold. Each subplot corresponds to a different meridional jet shift (s), while the color shading indicates the total number of hours during which the primary stationary anticyclone exists in each simulation. Within each subplot, the horizontal axis varies jet width (n), and the vertical axis varies jet vertical depth (b).

**4 Summary and Discussion**

This study has utilized the FV3-based idealized modeling framework to systematically disentangle the impacts of jet stream geometry on synoptic-scale extremes. By bridging the gap between classical baroclinic theory and modern operational dynamical cores, we have identified a set of robust "dynamical lookup tables" that map the latitude, width, and depth of the background flow to specific nonlinear outcomes. Our central finding is that the wave intensification, energetics, and the nonlinear saturation phase, specifically the occurrence of RWB, cyclone merging, and stationary anticyclones, is critically regulated by the width and structure of the jet.

Jet latitude, width, and altitude jointly control the baroclinic life cycle: poleward-shifted, broader (smaller n) and higher (larger b) jets accelerate early wave development, advance EKE peaks and promote more anticyclonic RWB events. These configurations also precondition the flow for earlier and more intensive cyclone merging, with larger transient near-surface wind maxima around coalescence. Stationary anticyclones emerge preferentially under poleward-shifted, broader, higher jets and extend persistence relative to equatorward, narrow, low-jet configurations.

Our experiments reveal that the "tug-of-war" between different jet configurations creates distinct regimes of extreme weather.

- Cyclone Merging Regime: This occurs when broad, poleward-shifted jets drastically reduce the background strain rate, creating a "kinematic sanctuary" that allows binary vortices to coalesce rather than being sheared apart. These merging events contribute to intense surface wind speeds, driven by the conservation of angular momentum during vortex capture.

- Stationary Anticyclone Regime: Conversely, we demonstrated that deep and broad jets dynamically precondition the atmosphere for persistent surface anticyclones. By reducing the effective planetary vorticity gradient and supporting larger zonal wavelengths, these configurations slow the intrinsic phase speed of Rossby waves to near-zero.

These findings provide a cohesive dynamical context for existing literature. The sensitivity of the wave-breaking lifecycle (LC1 vs. LC2) to jet latitude in our simulations successfully reproduces the paradigms established by Thorncroft et al. (1993). Furthermore, our results regarding jet width may offer a mechanistic interpretation for regional differences in cyclone activity. For example, the mid-winter Pacific jet is often in narrower conditions, which our simulations suggest should suppress cyclone merging and EKE. The finding appears qualitatively consistent with previous studies using two-layer quasi-geostrophic models, which suggest the "barotropic governor" associated with the narrower jet inhibits the growth of eddies (James, 1987; Harnik and Chang, 2004; Deng and Mak, 2005).

While this study relies on a dry, idealized physics suite, the exclusion of moisture and topography serves to isolate the baseline dynamical response of the atmosphere. By establishing the impacts of jet latitude and structure, we provide initial evidence towards a framework that helps interpret a wide range of observational and simulation results.(Zhang et al., 2021; Zilibotti et al., 2026) For instance, if future warming leads to a deepening of the eddy-driven jet, our results imply a potential shift toward more frequent cyclone merging and wind extremes. Similarly, a poleward shift in the jet stream may inherently lengthen the duration of stationary heat domes, a dynamic consistent with recent findings linking poleward-shifted weather regimes to anticyclonic wave breaking and clocking (Tamarin-Brodsky et al., 2026).

Naturally, direct extrapolation to the real world requires caution. The idealized nature of our experiments, such as relatively low model resolution, simplified single-jet structures, and the absence of latent heat release, means we likely underestimate the absolute intensity of merging and the duration of stationary anticyclones. Future work should address these simplifications by exploring broader parameter spaces, incorporating moisture effects, and testing multi-jet scenarios. Ultimately, this work underscores that monitoring the *shape* of the jet stream—not just its position—is essential for predicting the changing character of midlatitude weather extremes.

**Acknowledgments**

G.Z. thanks Profs. William R. Boos and Paul O'Gorman for stimulating discussions at Rossbypalooza at the University of Chicago. The research is supported by the U.S. National Science Foundation awards AGS-2327959 and RISE-2530555, as well as the faculty development fund of the University of Illinois Urbana-Champaign (UIUC). The authors would like to acknowledge the NSF National Center for Supercomputing Applications (NCSA) at UIUC for providing the high-performance computing resources and technical support that made the numerical simulations in this study possible.

**Open Research**

This project is committed to reproducible and accessible science by providing all necessary computational tools and source code. Standalone SHiELD Docker images are available via https://hub.docker.com/r/gfdlfv3/shield/tags to ensure a consistent model environment across different platforms, and the complete Python-based Initial Condition (IC) generation workflow can be accessed through the GitHub repository at https://github.com/xfr-123/Py-generated-IC-SHiELD.

**Conflict of Interest Disclosure**

The authors declare no conflicts of interest relevant to this study.

**Appendix A: Visualizations of different wind shear with different shear-related parameters**

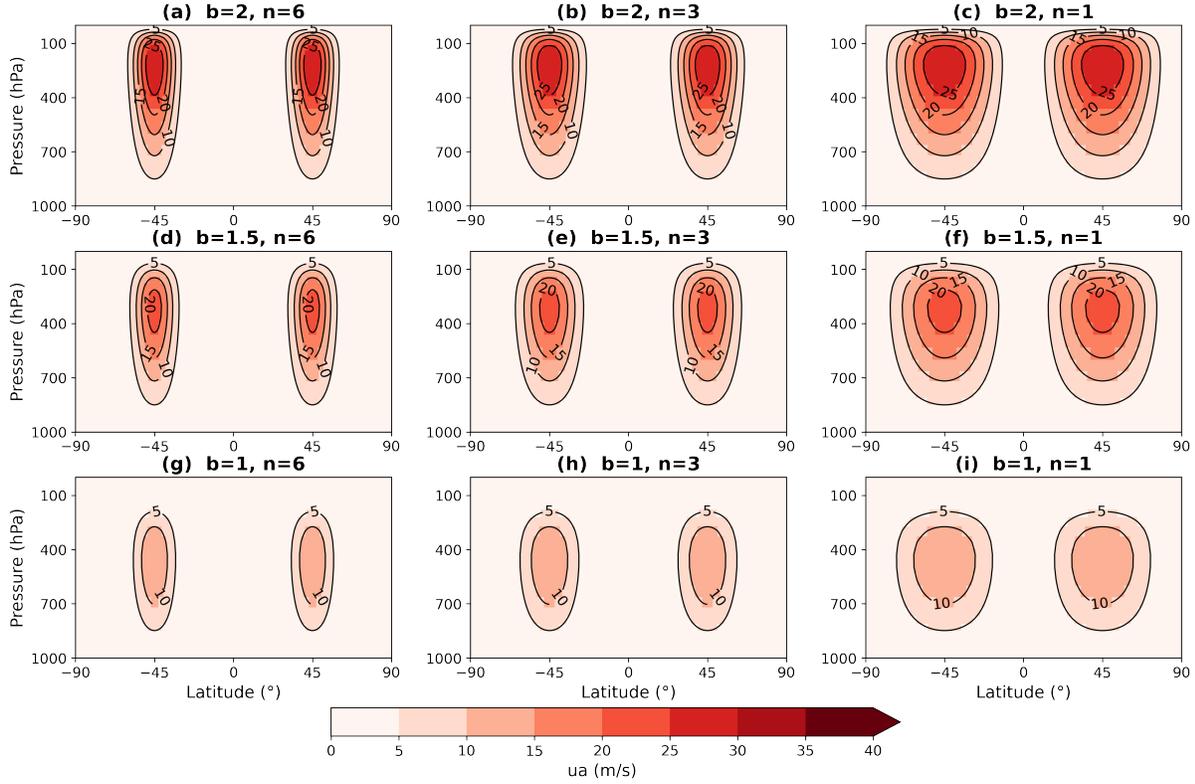

Figure 14. Variation of jet width and height parameters. Columns: A series of zonal wind profiles for different meridional widths (parameter *n*), illustrating a narrow vs. broad jet. Rows: Profiles for different vertical positions (parameter *b*), showing a low-altitude jet (blue) versus a high-altitude jet (orange). Wider jets (lower *n*) have gentler wind gradients, and high *b* shifts the wind maximum upward.

**Appendix B: Normalization of the Geopotential Field**

The gradient-wind balance equation provides the meridional derivative of the geopotential perturbation $\Phi'$. Because this defines $\Phi'$ only up to an arbitrary constant of integration, a closure condition is required to uniquely determine the field at each vertical level $\eta$.

To ensure that the perturbation represents a purely internal redistribution of mass and does not alter the global-mean geopotential defined by the reference state $\langle \Phi \rangle$, we impose a zero-mean constraint over the sphere:

$$\frac{1}{4\pi} \int_0^{2\pi} \int_{-\frac{\pi}{2}}^{\frac{\pi}{2}} \Phi'(\lambda, \phi, \eta) \cos\phi \, d\phi d\lambda = 0$$

In our numerical implementation, this is achieved by first integrating the gradient-wind balance equation from the equator to the poles to find a raw perturbation field. We then calculate the area-weighted global mean of that field and subtract it from all grid points. This step ensures that the initial state is energetically consistent and that the total mass of the atmosphere remains anchored to the specified reference profile.

Furthermore, by ensuring $\Phi'$ is zonally symmetric (independent of longitude $\lambda$) during this initialization, the zonal derivative $\frac{\partial \Phi'}{\partial \lambda}$ vanishes, thereby ensuring the initial geostrophic meridional wind is zero and preventing spurious inertia-gravity waves at the model start.

**Appendix C: Stability test of the idealized base state and the model**

Before introducing any perturbation to spawn baroclinic waves, it is crucial to verify that the generated initial conditions (ICs) are numerically and meteorologically stable. In other words, if the model is run with the IC but without the intentional perturbation that triggers instability, the flow should remain largely steady for an extended period. This test ensures that the initial state is properly balanced and that any subsequent wave growth is due to the intended perturbation, not an artifact of initial imbalance.

As an objective metric of stability, we computed the root-mean-square error (RMSE) of the zonal wind field at each time step, relative to the initial state, across all grid points and levels. Jablonowski & Williamson (2006) define a similar metric for dynamical core tests.

- $RMSE[\bar{u}(\lambda, \phi, \eta, t) - \bar{u}(\lambda, \phi, \eta, t = 0)] =$
$\left[\frac{1}{2} \int_0^1 \int_{-\frac{\pi}{2}}^{\frac{\pi}{2}} [\bar{u}(\lambda, \phi, \eta, t) - \bar{u}(\lambda, \phi, \eta, t = 0)]^2 \cos \phi \, d\lambda d\phi d\eta\right]^{\frac{1}{2}}.$

When $t = 0$, this RMSE is zero. If the flow remains steady, the RMSE should stay near zero; any growing deviation indicates drift or instability. In all our unperturbed runs, we found that the flow remained essentially stationary: there were no evident growing modes or oscillations. Quantitatively, the RMSE of zonal wind after 15 days was on the order of $10^0$ (a few m/s in units) for all cases, effectively zero, given the model precision. For reference, the "default" baroclinic wave test case encoded with FV3 (a standard midlatitude jet) also exhibits an RMSE on the order of $10^0$ over 15 days, so our generated cases are comparably stable.

These stability tests build confidence that our ICs are correctly balanced. The shifted-jet cases, which were not part of previous standard tests, also showed no pathological behavior, indicating that our analytical solution and balancing procedure remained valid even when the jet is displaced far from its usual latitude.